\documentclass[10pt,a4paper]{article}            
\usepackage{geometry}                
\usepackage{fancyhdr}
\usepackage{authblk}
\usepackage{graphicx}
\usepackage{url}
\makeatletter
\def\url@leostyle{%
  \@ifundefined{selectfont}{\def\UrlFont{\sf}}{\def\UrlFont{\small\ttfamily}}}
\makeatother
\usepackage{epstopdf}
\usepackage{amssymb}
\usepackage[fleqn]{amsmath}
\def\mathbi#1{\textbf{\em #1}}
\numberwithin{equation}{section}
\DeclareMathSymbol{\Gamma}{\mathalpha}{letters}{"00}
\DeclareMathSymbol{\Lambda}{\mathalpha}{letters}{"03}
\DeclareMathSymbol{\Omega}{\mathalpha}{letters}{"0A}
\DeclareMathAlphabet{\mathitbf}{OML}{cmm}{b}{it}

\begin{document}                  



\title{The Geometry of Niggli Reduction II: BGAOL -- Embedding Niggli Reduction }


\author[1,*]{Lawrence C. Andrews }
\affil[1]{Micro Encoder  Inc., 11533 NE 118th St, \#200,  Kirkland, WA 98034-7111 USA}
\author[2,*]{Herbert J. Bernstein }
\affil[2]{Dowling College, 1300 William Floyd Parkway, Shirley, NY 11967 USA}
\affil[*]{To whom correspondence should be addressed. Email: {\it yaya@dowling.edu}}









\maketitle                        

\begin{abstract}
Niggli reduction can be viewed as a series of operations in a \mbox{six-dimensional} space
derived from the metric tensor.  An implicit embedding of the space of Niggli-reduced cells in
a higher dimensional space to facilitate calculation of distances between cells is described.
This distance metric is used to create a program, BGAOL, for Bravais lattice determination.
Results from BGAOL are compared to the results from other metric-based Bravais lattice determination
algorithms.
\end{abstract}


\section{Introduction}

BGAOL (Bravais General Analysis of Lattices) is a Bravais lattice identification program based on
the ${\mathbi{G}^{\mathbi{6}}}$ analysis of Niggli reduction described in the first paper of this series \cite{Andrews2012}.
Niggli reduction defines a complex space that has not previously been fully analyzed. 
Several authors have published interesting commentaries on the properties \cite{Hosoya2000} \cite{Oishi-Tomiyasu2012}  \cite{Gruber1997}. These studies use the space ${\mathbi{G}^{\mathbi{6}}}$ \cite{Andrews1988}, or a similar metric-tensor-based space, or 
a projection of ${\mathbi{G}^{\mathbi{6}}}$ to a space of lower dimensionality, respectively. Two principal uses of Niggli reduction are 
the determination of Bravais lattice type and the construction of databases using a representation of the unit cell for the key \cite{Andrews1980} \cite{Toby1994} \cite{Byram1996}.  

Both uses can be viewed as distance determinations in ${\mathbi{G}^{\mathbi{6}}}$. In the former case the distances to the 
Bravais lattice subspaces are used, and in the latter case the distances between pairs of unit cells
are used.  However, the complexity of the space has consequences in some regions; it is not adequate to consider only one representation of a unit cell in ${\mathbi{G}^{\mathbi{6}}}$.   A standard mathematical solution is 
to create an ``embedding'' \cite{Nash1956} of the space with an appropriate associated metric.
In such an embedding, separate regions of the space under consideration are sewn together into
a single fundamental region preserving distances from the original piecewise presentation.

In the case being considered the regions contain sets of cells that appear to be far apart
originally but which can be seen to represent similar lattices as the regions are sewn together
and sets of cells that remain far apart after the sewing can be seen not to represent similar lattices. 
In concept, this is similar to what we do in folding of atomic coordinates into the 
asymmetric unit of a crystal.  This is an example of a simple embedding, allowing us to see
which atoms interact. This is the approach followed in BGAOL.

In order to define an embedding, the operations defining the fundamental region must be specified.  
In the case of Niggli reduction, the complete space is ${\mathbi{G}^{\mathbi{6}}}$, and the fundamental region is the fraction of the space containing only Niggli-reduced cells. Proper unit cells in any other region of ${\mathbi{G}^{\mathbi{6}}}$ can be transformed into the fundamental region by the rules of Niggli reduction. The transformations at the boundaries must be enumerated and their combinations analyzed \cite{Andrews2012}.  See Table \ref{5D}.

Given the complete set of conditions that define all boundaries of the fundamental unit and their relationships to adjacent units, the transformations of coordinates on crossing the boundaries are enumerated.  \cite{Oishi-Tomiyasu2012} has enumerated transformations in a related space. 

\section{Background}

Crystallography began with the study of crystal morphology and the classification of substances
by the shapes of their crystals, a database concept before the creation of databases.  Von Laue 
\cite{vonLaue1967} provided an accessible
description.  In the present context, two themes have developed: Bravais lattice
assignment and database searches to identify substances by their unit cell parameters. 

\subsection{Bravais Lattice Assignment}

Modern work on Bravais lattice assignment has taken two directions:  qualitative absolute
assignment of lattice type versus quantitative assignment using a metric to measure the distance
from the 14 Bravais lattice types.   This is a fuzzy distinction because all methods are
fundamentally quantitative, being rooted in numeric cell parameters.  The advantage of
the methods based on, and making full use of, a metric is that they perform well in the
presence of experimental error.
The more fine-grained the metric used, the more easily and efficiently can the alternatives
be ranked.  Gruber's work \cite{Gruber1997} is the latest in qualitative assignment of
lattice types.  DELOS \cite{Zimmermann1985} is a popular example of a rather coarse-grained
metric.  Kabsch has incorporated a fine-grained metric in XDS \cite{Kabsch1993} \cite{Kabsch2010} based
on the sum of the magnitudes of deviations from the various Niggli reduction conditions.  See \cite{Macicek1992} for a reasonably complete review of the relevant literature.

The use of a fine-grained metric under which it is meaningful to ask precisely how far a probe
cell is from a given lattice and to compare that distance to the experimental error
began with \cite{Andrews1988}, in which the space ${\mathbi{G}^{\mathbi{6}}}$, consisting
of vectors
\[ 
\overrightarrow{g} =  \left[ \overrightarrow{a}. \overrightarrow{a}, \overrightarrow{b}. \overrightarrow{b}, \overrightarrow{c}. \overrightarrow{c}, 2\overrightarrow{b}. \overrightarrow{c}, 2\overrightarrow{a}. \overrightarrow{c}, 2\overrightarrow{a}. \overrightarrow{b}\right]
\]
\[
\text{~~~}= [a^2, b^2, c^2, 2 b c\text{~} cos \alpha, 2 a c\text{~} cos \beta, 2 a b\text{~} cos \gamma ]
\]
  (a modified metric tensor), was introduced.
The concept is simple, but the implementation is complex because a very large number of 
iterations may be necessary to apply the boundary transformations of the Niggli cone.  BLAF
\cite{Macicek1992}  and OT-BLD \cite{Oishi-Tomiyasu2012} cut off the iterations, creating the possibility
of missed symmetries.  The implementation of 
\cite{Andrews1988}, ITERATE, continues without a cutoff until no new candidates are found, to
avoid missed symmetries but at the expense of additional execution time.  BGAOL resolves
this conflict by specifically using the 15 5-D boundaries cited in \cite{Andrews2012} labelled {\it 1, 2, 3, 4, 5, 6, 7, 8, 9, A, B, C, D, E, F},
from which the many remaining internal boundaries of the Niggli cone may be derived as intersections, 
and by using an isometric embedding that sharply limits the boundary transforms to be applied
to the ones directly involved with those 15 5-D boundaries, plus three of the 4-D boundaries 
({\it 8F, BF} and {\it EF}) in the
negative portion of the Niggli cone and two boundaries ({\it 69} and {\it 6C}) in the positive portion of the
Niggli cone that contains the images of {\it 8F}, {\it BF} and {\it EF} under their boundary transforms.

\subsection{Database Searches}

For Niggli reduced cells, the last three elements of the ${\mathbi{G}^{\mathbi{6}}}$ vector are
highly unstable under small perturbations of the cell parameters \cite{Andrews1980}.  That is why
many iterations have been needed in robust lattice identification. 
The  highly iterative nature of prior uses of the ${\mathbi{G}^{\mathbi{6}}}$ metric along with a lack
of clearly defined stopping criteria creates
a significant burden for application to cell databases.  One feasible approach
with an easier-to-compute metric has been to use perturbation-stable
subsets of the cell parameters, such as the edge lengths and the volume, as
in the JCDPS database originally distributed on cards in the 1930s \cite{Wong-Ng1982}.  This approach proved insufficiently selective as the number of solved
structures grew.  \cite{Andrews1980} added the reduced reciprocal cell edge lengths to
the search key, and that approach was also successfully implemented by 
\cite {Bernstein1979} \cite{NIH/EPA1980} \cite{Rodgers1992} \cite{Byram1996} \cite{Toby1994}.  This metric, however,
is very non-linear in comparison to the ${\mathbi{G}^{\mathbi{6}}}$ metric.  A database search 
based on the combinatorial approach in \cite{Andrews1988} is used in WebCSD \cite{Thomas2010}.
A ``Nearest Cell'' search based on 3-D unit cell vectors can be found in \cite{Ramraj2011}.
The ${\mathbi{G}^{\mathbi{6}}}$ embedding can have a significant impact on such searches.
In addition, it is helpful in the context of databases to linearize and scale the
{\AA}ngstroms squared distance of ${\mathbi{G}^{\mathbi{6}}}$ to linear {\AA}ngstroms.
This will be discussed in a subsequent paper in this series \cite{McGill2013}.

\subsection{Embeddings}

The problem of finding how far cells in the Niggli cone are from other cells in the Niggli
cone is similar to the problem of finding the distance between atoms in
a crystal, where the shortest distance may not be the distance within the asymmetric
unit of the chosen cell.  For example in Fig. \ref{fig:Cell_fig}, two atoms, A and B,  are shown in
the asymmetric unit of a cell chosen from a \mbox{two-dimensional} lattice, with symmetry-related
copies of those atoms in neighboring cells.  In this case, a symmetry-related copy of B
is closer to A (shown with a solid black line) than is the original B in the same asymmetric
unit.  An alternative to searching through neighboring cells in two dimensions would be to
pick up the matching left and right edges of the cell and glue them together to form a tube,
and then bend the tube to glue the remaining edges together to form a torus.  Then
we can navigate between points on the surface of the torus looking for the shortest
distance.  In that representation of this cell, the shortest path from A to B
is the three pieces shown in Fig. \ref{fig:Cell_fig} as the bold black segments 1, 2 and 3.
Even though they appear to be disjoint in the \mbox{two-dimensional} representation, they are contiguous
in the embedding.

This process of picking up a lower-dimensional manifold in which
we know the geometry in Euclidean patches and gluing the edges of the patches together
to form a closed surface in a higher-dimensional space but with the same distances between 
points is called an isometric embedding.  ``All'' we need to know is the distance function on
that embedded surface.

Embedding the Niggli cone in ${\mathbi{G}^{\mathbi{6}}}$ into a higher-dimensional space is,
as one might expect, more complicated than embedding a \mbox{three-dimensional} lattice
as a torus-like object into a \mbox{six-dimensional} space.  In addition to having more dimensions,
the symmetry operations generated by the boundary transformations in Table \ref{5D}
are not, in general, isometric.  The face-diagonal and body-diagonal boundary transformations significantly
compress space in some directions and expand it in others.  In measuring the
distance between cells in the Niggli cone, we always have to measure distances between
representatives of cells within the Niggli cone and not even between a representative in the cone and one outside.
We can, however, safely ``unroll'' the cone into multiple images using the equal-cell-edge and
ninety-degree boundary transforms and then measure distances among those cell representatives 
because the associated transforms for those boundaries are isometric.  The non-isometric
boundary transforms have anisotropic expansions and contractions, ranging from an expansion
by a factor of nearly 3.6 in one direction and a contraction by a factor of less than 0.28 in
another direction for the ${\mathbi{G}^{\mathbi{6}}}$ vectors 
(corresponding to an expansion factor of nearly 1.9
and a contraction factor of less than 0.53 for the ${\mathbi{R}^{\mathbi{3}}}$ cells) for the 
body-diagonal boundary transform.  For the face-diagonal boundaries the corresponding
expansions and contractions are 2.8 and .36 for  the ${\mathbi{G}^{\mathbi{6}}}$ vectors 
(corresponding to expansions and contractions of nearly 1.7 and less than 0.6 for the
${\mathbi{R}^{\mathbi{3}}}$ cells).  The face-diagonal and body-diagonal boundary transforms 
act like folds around special position subspaces of the boundary manifolds, forming cones 
rather than tori in an embedding, with no impact from the metric distortions on the
boundary manifolds themselves.  However, if we step past those boundaries into the
surrounding ${\mathbi{G}^{\mathbi{6}}}$ environment, the effect of those metric
distortions is very much like looking through glass of a high anisotropic refractive index thereby potentially
creating a very large number of distorted images of the distances within the Niggli
cone.  Using the embedding and confining distance
measurements to those entirely within the cone greatly reduces this computationally
expensive effect and the need for inappropriately early terminations of iterations.

\section{The BGAOL Embedding Distance}

There are two ways in which to compute an embedded distance.  In the first way, one
maps the lower-dimensional space into a higher-dimensional space and computes
distances along the resulting curved surface using the coordinate system
of the higher-dimensional space, much as one computes spherical
distances on the surface of the earth to determine the distance between 
cities \cite{ASocietyofGentlemen1771}.
In the second way, one uses the coordinate system of the lower-dimensional
space and computes distances in those terms, using the rules of the embedding
to join patches together \cite{Helgason1962}.  Both approaches can involve comparisons among multiple
alternate distances, just as one might have to compare going east versus going west
in deciding on the shortest distance between New York, USA and Sydney, Australia.
In BGAOL, we chose to work with the coordinate system in ${\mathbi{G}^{\mathbi{6}}}$
rather than with curvilinear coordinates in a higher-dimensional space.

The program BGAOL computes 
the embedded distance between ${\mathbi{G}^{\mathbi{6}}}$ vectors $v_1$ and $v_2$,
which must both be within the Niggli cone.  This restriction is important because
the boundary transformations are not isometric and have significant anisotropies, causing
the regions outside the Niggli cone to be viewed as if through glass of anisoropic
refractive index.  The distances are computed 
from the  ${\mathbi{G}^{\mathbi{6}}}$ coordinates as follows.

\begin{itemize}
\item{1.  Unroll the Niggli cone by applying the six permutations resulting from
interchanging the cell edges and the four possible acute-obtuse angle changes,
for an initial set of 24 alternate presentations of each cell, $v$:
$v$, $M_1 v_1$, $M_2 v_1$, $M_1 M_2 v_1$, $M_2 M_1 v_1$, $M_2 M_1 M_2 v_1$,
$M_3 v$, $M_3 M_1 v_1$, $M_3 M_2 v_1$, $M_3 M_1 M_2 v_1$, $M_3 M_2 M_1 v_1$, $M_3 M_2 M_1 M_2 v_1$,
$M_4 v$, $M_4 M_1 v_1$, $M_4 M_2 v_1$, $M_4 M_1 M_2 v_1$, $M_4 M_2 M_1 v_1$, $M_4 M_2 M_1 M_2 v_1$,
$M_5 v$, $M_5 M_1 v_1$, $M_5 M_2 v_1$, $M_5 M_1 M_2 v_1$, $M_5 M_2 M_1 v_1$, $M_5 M_2 M_1 M_2 v_1$.}

\item{2.  For each of the 48 resulting cells from step 1, 
compute the distances to and projections onto to each of the 15 5-D
Niggli cone boundaries.}

\item{3.  For each of the 48 resulting cells from step 1,
compute the distances to and projections onto each of the 
three intersections between the face-diagonal cases and
the body-diagonal case in the negative (obtuse angle) portion
of the Niggli cone: {\it 8F, BF, EF} as well as the
distances to and boundary mapping onto the images of those
intersections in the positive (acute angle) portion
of those intersections.  Specifically, for each cell, $v$,  compute
the distance, projections and images: \\
$||(I-P_{8F}) v||$, $P_{8F} v$,  $M_8 P_{8F} v$ \\
$||(I-P_{BF}) v||$, $P_{BF} v$,  $M_B P_{BF} v$ \\
$||(I-P_{EF}) v||$, $P_{EF} v$,  $M_E P_{EF} v$ \\
$||(I-P_{6C}) v||$, $P_{6C} v$,  $M_C P_{6C} v$ \\
$||(I-P_{69}) v||$, $P_{69} v$,  $M_6 P_{69} v$ and $M_9 P_{69} v$ \\}

\item{4.  For
all subsequent distance calculations, also unroll the Niggli standard form transformations
that restrict the Niggli cone to $+++$ or $- - -$, by defining
\[
dist_{456}( x , y) = min( || x - y ||,
\]
\[
\text{~~~~~~~~~~} || [x_1,x_2,x_3,x_4,x_5,x_6] - [y_1,y_2,y_3,y_4,-y_5,-y_6]|| ,
\]
\[
\text{~~~~~~~~~~} || [x_1,x_2,x_3,x_4,x_5,x_6] - [y_1,y_2,y_3,-y_4,y_5,-y_6]|| ,
\]
\[
\text{~~~~~~~~~~} || [x_1,x_2,x_3,x_4,x_5,x_6] - [y_1,y_2,y_3,-y_4,-y_5,y_6]|| )
\]
}
\item{5.  Compute the direct minimum distance from each of the 24 images of the first
cell from step 1 to each of the 24 images of the second cell from step 1}.
\item{6.  Then for each of the 15 5-D boundaries and for each of the 
576 combinations of one of the 24 images of the first cell and one of the
24 images of the second cell, compute the minimum of the distances computed
thus far and the distance going from the first cell to the chosen boundary
and then from the boundary to the second cell, treating each projection into
a boundary and its transformation using the boundary transformation as
equivalent.}
\item{7. For each member of each set of permutations, compute the distance from each
permutation to each of the face-diagonal and body-diagonal boundary manifolds.  The
face-diagonal boundaries are grouped together as three cases ({\it 6-7-8, 9-A-B, C-D-E}), with two subcases each.  In the first three cases these are 
the full \mbox{five-dimensional} boundaries, and in the subcases these are the \mbox{four-dimensional} boundaries produced
by the intersections with the body-diagonal boundary manifold ({\it 8F, BF, EF}).}
\item{8. For each face-diagonal or body-diagonal boundary manifold, $\Gamma$, consider
a member $w_1$ from the first set of permutations and $w_2$ from the
second set of permutations.  Let $h_1$ be the distance from $w_1$ to $\Gamma$
and $h_2$ be the distance from $w_2$ to $\Gamma$.}
\item{9. If $h_1 + h_2$ is less than the minimum distance already found, let $P_{\Gamma} w_1$
be the projection of $w_1$ onto $\Gamma$ and $M_{\Gamma} P_{\Gamma} w_1$ be the
image of that projection under the boundary transformation and let $P_{\Gamma} w_2$
be the projection of $w_2$ onto $\Gamma$ and $M_{\Gamma} P_{\Gamma} w_2$ be the
image of that projection under the boundary transformation.  For the {\it 8F, BF} and {\it EF} 
\mbox{four-dimensional} boundaries, use the transformations for the corresponding
face-diagonal boundaries ($M_8, M_F, M_E$).  Compare the minimum
distance thus far to 
\[
\sqrt{((h_1+h_2)^2+ min (dist_{456}(P_{\Gamma} w_1, P_{\Gamma} w_2),}
\]
\[
\text{~~~~~~~~~~}  \overline{dist_{456}(P_{\Gamma} w_1, M_{\Gamma}P_{\Gamma} w_2),}
\]
\[
\text{~~~~~~~~~~} \overline{dist_{456}(M_{\Gamma}P_{\Gamma} w_1, P_{\Gamma} w_2),}
\]
\[
\text{~~~~~~~~~~}  \overline{dist_{456}(M_{\Gamma}P_{\Gamma} w_1, M_{\Gamma}P_{\Gamma} w_2))^2)}
\]
}

and keep the smaller value.

The raw distance in ${\mathbi{G}^{\mathbi{6}}}$ is not sufficient for comparison of
lattices of different symmetries and does not consider distances in relationship
to the size of experimental errors.  The anorthic lattices have the full six degrees of freedom
of the space, monoclinic have four, orthorhombic have three, hexagonal and tetragonal have
two and cubic have one.  Multiplying the reported ${\mathbi{G}^{\mathbi{6}}}$ distances
by the square root of the number of degrees of freedom provides a better comparision
among possible lattices.  If we then divide by the ${\mathbi{G}^{\mathbi{6}}}$ experimental
error estimate, we get a dimensionless ``Z-score''.

\end{itemize}

Computationally, the multiplicities of the combinations used is lower than
one might expect because of constant pruning by comparing the distance
computed at each stage to the distance to the boundary under consideration.
If the boundary distance, which was precomputed in step 2 or 3 is larger than the
previously computed minimum distance between cells, there is no need to compute 
path lengths that include that boundary distance.

\goodbreak

\section{Implementation of the embedding}

BGAOL is a modification of our earlier, iteration-based program ITERATE \cite{Andrews1988}
using embedding-based distances to search for likely Bravais lattice matches.  
The only other lattice matching programs that we know of that use a metric are 
BLAF \cite{Macicek1992}, DELOS \cite{Zimmermann1985}, XDS \cite{Kabsch2010} and OT-BLD, the lattice matching part of CONOGRAPH \cite{Oishi-Tomiyasu2012}.  BLAF uses an $L_1$ measure on the metric tensor, while ITERATE and BGAOL use an $L_2$ measure. DELOS uses a coarse measure based on ``cycles''. OT-BLD reports matches using a fractional measure, also based on the metric tensor.  XDS uses a
``quality index'' based on the sum of the extents to which the inequalities of Niggli
reduction are not satisfied for components of the metric tensor, essentially an $L_1$ measure
of the distance from each Niggli-cone boundary polytope.   Table \ref{tab:comparison1}  shows a comparison of BGAOL results to other programs, except XDS.  Table \ref{tab:comparison1}  shows a comparison
of the BGAOL Z-score to the XDS quality indicator.

\subsection{Distance calculation}
\begin{figure}
\caption{To illustrate distance calculations between arbitrary points, a line was drawn in ${\mathbi{G}^{\mathbi{6}}}$ between an unreduced point close to $cF$ and its reduced image.  Each curve shows the distances from 100 points along each line to the corresponding reduced form. The upper curve starts from $a=3.162, b=3.173, c=3.163, \alpha=60.094, \beta=60.049, \gamma=60.338$ and ends at $a=3.162, b=3.163, c=3.173, \alpha=60.115, \beta=89.843, \gamma=60.049$. The lower curve starts from $a=3.171, b=3.166, c=3.160, \alpha=60.265, \beta=59.999, \gamma=60.161$ and ends at $a=3.160, b=3.165, c=3.166, \alpha=90.190, \beta=119.735, \gamma=119.825$.
}
\scalebox{.45}{\includegraphics{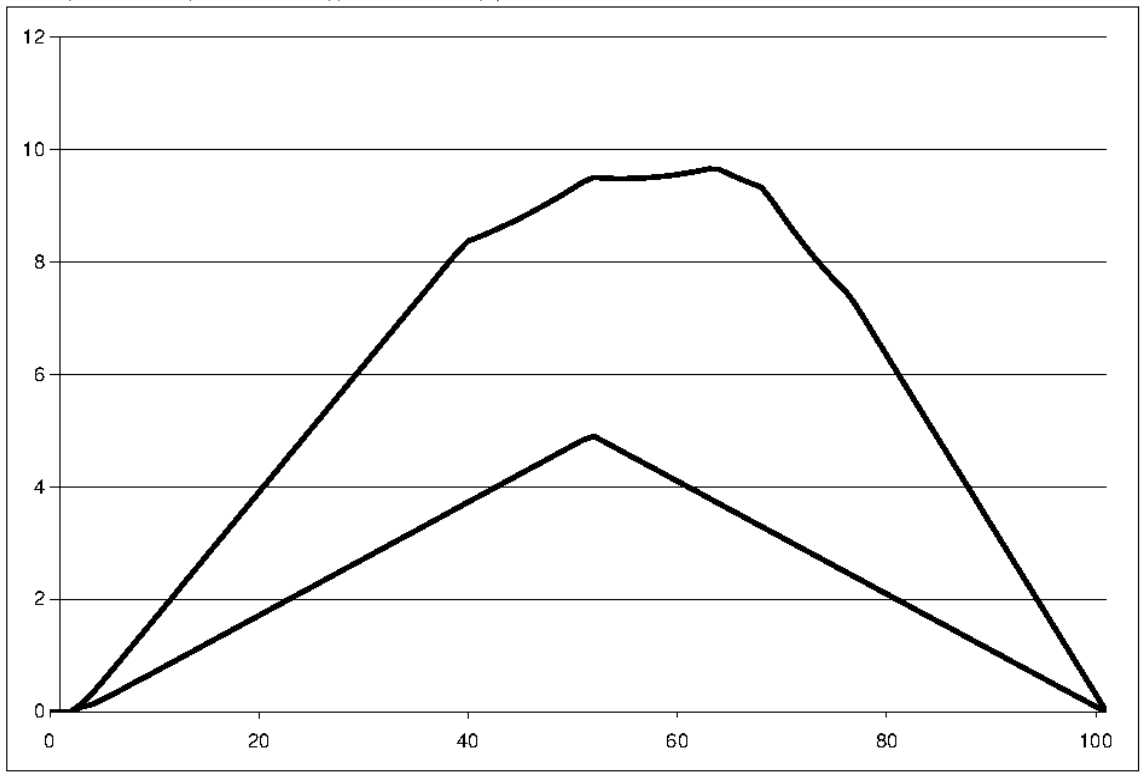}} 
\label{fig:Follower}
\end{figure}

BGAOL distances are calculated using the function NCDIST, which computes the distance between pairs of reduced cells. Bravais lattice determination for a given probe cell consists of finding which boundary polytopes of the Niggli cone are closest to the probe. Constructing and using a cell database requires computing the distance between cells as points in ${\mathbi{G}^{\mathbi{6}}}$ that are arbitrarily far apart.  Figure \ref{fig:Follower} illustrates the use of NCDIST to compute distances between well-separated points.

\subsection{Availability and Test Results}
A BGAOL-based lattice identification web server is available at 

\noindent{}\verb|http://www.bernstein-plus-sons/software/BGAOL/|

\noindent{}A source kit may also be downloaded from a link on that page.

The prior ITERATE-based lattice identification web server is
available at 

\noindent{}\verb|http://www.bernstein-plus-sons.com/software/ITERATE/|

\noindent{}The latest version of the source code
of BGAOL is maintained on SourceForge for svn access at 

\noindent{}\verb|svn checkout| 

\noindent{}\verb|  svn://svn.code.sf.net/p/iterate/code/trunk/bgaol|

\noindent{}\verb|    bgaol-code|.

\noindent{}The source kit contains the test program, Follower.for, that computes the distance for database work as shown in Fig. \ref{fig:Follower}. 

The database code that will be discussed in a subsequent paper is available
from the ``sauc'' module in the same repository.



\section*{Acknowledgements}

The authors acknowledge the invaluable assistance of Frances C. Bernstein.
\\
The work by Herbert J. Bernstein has been supported in part by NIH NIGMS grant GM078077.
The content is solely the responsibility of the authors and does not
necessarily represent the official views of the funding agency.
\\
Lawrence C. Andrews would like to thank Frances and Herbert Bernstein for hosting him during 
hurricane Sandy and its aftermath. Elizabeth Kincaid has contributed significant support in many ways.


\onecolumn

\begin{table}
\caption{Fifteen 5-D boundary polytopes of Niggli-reduced cells in ${\mathbi{G}^{\mathbi{6}}}$.   
For a given boundary polytope ${\mathitbf{\Gamma}}$, the
column ``Condition'' gives the ${\mathbi{G}^{\mathbi{6}}}$ constraints (prior to closure) of the
boundary polytope.   Boundary polytopes {\it 1} and {\it 2}
apply in both the all acute ($+ + +$) and all obtuse  ($- - -$) branches of the Niggli-reduced
cone.  Boundary polytopes {\it 8, B, E} and {\it F} are restricted to the all obtuse ($- - -$) branch of the
Niggli-reduced cone, $\mathitbf{N}$.  Boundary polytopes {\it 6, 7, 9, A, C} and {\it D}  are restricted to the all acute ($+ + +$) branch of $\mathitbf{N}$.
Boundary polytopes {\it 3, 4} and {\it 5} are boundaries of both the all acute ($+ + +$) and all obtuse  ($- - -$) branches.
}
\begin{center}
\begin{tabular}{llll}
{\bf Class}& {\bf Boundary}& {\bf Condition}&{\bf Transformation Matrix}\\
\multicolumn{2}{l}{Equal cell edges}&&\\
&{\it 1} &$g_1 = g_2$&$\left[010000/100000/001000/000010/000100/000001\right]$\\
&{\it 2} &$g_2 = g_3$&$\left[100000/001000/010000/000100/000001/000010\right]$\\
\multicolumn{2}{l}{Ninety degrees}&&\\
&{\it 3} &$g_4 = 0$&$\left[100000/010000/001000/000100/0000\overline{1}0/00000\overline{1}\right]$\\
&{\it 4} &$g_5 = 0$&$\left[100000/010000/001000/000\overline{1}00/000010/00000\overline{1}\right]$\\
&{\it 5} &$ g_6 = 0$&$\left[100000/010000/001000/000\overline{1}00/0000\overline{1}0/000001\right]$\\
\multicolumn{2}{l}{Face diagonal}&&\\
&{\it 6} &$g_2 = g_4 $ and $ g_5 \geqslant  g_6$&$\left[100000/010000/011\overline{1}00/0\overline{2}0100/0000\overline{1}1/00000\overline{1}\right]$\\
&{\it 7} &$g_2 = g_4 $ and $ g_5 <  g_6$&$\left[100000/010000/011\overline{1}00/020\overline{1}00/0000\overline{1}1/000001\right]$\\
&{\it 8} &$g_2 = -g_4$&$\left[100000/010000/011100/020100/0000\overline{1}\overline{1}/00000\overline{1}\right]$\\
&{\it 9} &$g_1 = g_5 $ and $ g_4 \geqslant  g_6$&$\left[100000/010000/1010\overline{1}0/000\overline{1}01/\overline{2}00010/00000\overline{1}\right]$\\
&{\it A} &$g_1 = g_5 $ and $ g_4 < g_6$&$\left[100000/010000/1010\overline{1}0/000\overline{1}01/2000\overline{1}0/000001\right]$\\
&{\it B} &$g_1 = -g_5$&$\left[100000/010000/101010/000\overline{1}0\overline{1}/200010/00000\overline{1}\right]$\\
&{\it C} &$g_1 =  g_6 $ and $ g_4 \geqslant g_5$&$\left[100000/11000\overline{1}/001000/000\overline{1}10/0000\overline{1}0/\overline{2}00001\right]$\\
&{\it D} &$g_1 =  g_6 $ and $ g_4 < g_5$&$\left[100000/11000\overline{1}/001000/000\overline{1}10/0000\overline{1}0/20000\overline{1}\right]$\\
&{\it E} &$g_1 = - g_6$&$\left[100000/110001/001000/000\overline{1}\overline{1}0/0000\overline{1}0/200001\right]$\\
\multicolumn{2}{l}{Body diagonal }&&\\
&{\it F} &$g_1+g_2+g_4+g_5+ g_6 = 0$&
$\left[100000/010001/111111/0\overline{2}0\overline{1}0\overline{1}/\overline{2}000\overline{1}\overline{1}/000001\right]$\\
\end{tabular}
\end{center}
\label{5D}
\end{table}%

\begin{table}
\caption{Search results for BGAOL, BLAF, DELOS, OT-BLD and ITERATE for
basic beryllium acetate, $a=19.2600, b=63.8825, c=27.2394, \alpha=5.7696,
\beta=19.4709,\gamma=17.2952$ \cite{Himes1987}.  The BGAOL, BLAF and ITERATE
columns show distances in ${\AA}^2$ from the probe to the appropriate
boundary manifold for the indicated Niggli Lattice character in ${\mathbi{G}^{\mathbi{6}}}$.
The BLAF column uses $L_1$ distances versus $L_2$ distances for the others.
The DELOS column shows the numbers of cycles of relaxation of the Delaunay
reduction needed to find the indicated symmetry.  The OT-BLD column is
a dimensionless fractional measure on the agreement of metric tensors.
}
\begin{center}
\begin{tabular}{llllll}
{\bf Lattice}&&&&&\\
{\bf Character}& {\bf BGAOL}& {\bf BLAF}&{\bf DELOS}&{\bf OT-BLD}&{\bf ITERATE}\\
cF	&0.067	&	&cycle1	&1E-05	&0.096\\
tI	&0.038	&0.014	&cycle1	&6E-06	&0.038\\
hR	&0.065	&0.013	&cycle1	&7E-06	&0.076\\
oF	&0.038	&	&cycle1	&6E-06	&0.038\\
oI	&0.019	&0.007	&cycle1	&4E-06	&0.018\\
mI	&0.007	&0.004	&cycle1	&1E-06	&0.007\\
\end{tabular}
\end{center}
\label{tab:comparison1}
\end{table}%

\begin{table}
\caption{Partial search results for BGAOL and XDS for the test cell,
$a=62.1, b=63.5, c=92.9, \alpha=90.0, \beta=90.1, \gamma=107.200$ 
from \cite{Kabsch1993}.  The BGAOL column shows the 
${\mathbi{G}^{\mathbi{6}}}$ distance, ${\mathbi{G}^{\mathbi{6}}}$
degrees-of-freedom weighted Z-score, assuming errors of 0.2 in edges and 0.1 on angles,
which in this case corresponds to a ${\mathbi{G}^{\mathbi{6}}}$ error estimate of
61.3 $\AA ^2$.  The XDS QI (Quality indicator) and scaled QI columns shows the raw 
deviations from Niggli reduction and the same deviations scaled by 
$100/(0.2*min(a^2,b^2,c^2))$ and capped at 999 as per \cite{Kabsch2013}.  Only the three most promising lattice types are shown.  The structure
solution gave $oC$.}
\begin{center}
\begin{tabular}{lllll}
{\bf Lattice}&${\mathbi{G}^{\mathbi{6}}}$&{\bf BGAOL}&{\bf XDS}&{\bf Scaled}\\
{\bf Character}&{\bf Distance}&{\bf Z-Score}&{\bf QI}&{\bf QI}\\

mP&20.138&0.657&1.0&.13\\
oC&125.958&3.560&23.8&3.09\\
mC&125.150&4.085&23.4&3.03\\
\end{tabular}
\end{center}
\label{tab:comparison2}
\end{table}%


\begin{figure} 
\caption{Example of the difficulty of finding the shortest distance in a lattice.  The distance
within the asymmetric unit of the chosen cell is shown as an orange dotted line.  However,
the shortest distance, shown as a solid back line, crosses multiple unit cells. } 
\scalebox{.55}{\includegraphics{Cell_fig.epsf}} 
\label{fig:Cell_fig.espf}
\end{figure} 

\bibliographystyle{plain}
\bibliography{BGAOL_refs}

\end{document}